\documentclass[manuscript]{aastex}

\shorttitle{GEMINI observations of Liller 1}
\shortauthors{Saracino et al.}

\begin{document}

\title{GEMINI/GeMS observations unveil the structure of the heavily obscured globular
  cluster Liller 1\footnote{\footnotesize Based on observations 
  obtained at the Gemini Observatory, which is operated by the 
  Association of Universities for Research in Astronomy, Inc., 
  under a cooperative agreement with the NSF on behalf of the 
  Gemini partnership: the National Science Foundation (United States), 
  the National Research Council (Canada), CONICYT (Chile), the Australian 
  Research Council (Australia), Minist\'{e}rio da Ci\^{e}ncia, Tecnologia 
  e Inova\c{c}\~{a}o (Brazil) and Ministerio de Ciencia, Tecnolog\'{i}a e 
  Innovaci\'{o}n Productiva (Argentina).
Based on observations gathered with ESO-VISTA telescope (program ID 179.B-2002).}
}
\author{S. Saracino\altaffilmark{1,2}, E. Dalessandro\altaffilmark{1},
  F. R. Ferraro\altaffilmark{1}, B. Lanzoni\altaffilmark{1}, D. Geisler\altaffilmark{3},
  F. Mauro\altaffilmark{4,3}, S. Villanova\altaffilmark{3}, C. Moni
  Bidin\altaffilmark{5}, P. Miocchi\altaffilmark{1}, D. Massari\altaffilmark{1} 
  }

\altaffiltext{1}{Dipartimento di Fisica e Astronomia, Universit\`a di
  Bologna, Viale Berti Pichat 6/2, I-40127 Bologna, Italy}
\altaffiltext{2}{INAF - Osservatorio Astronomico di Bologna, via
  Ranzani 1, I-40127 Bologna, Italy}
\altaffiltext{3}{Departamento de Astronom\'ia, Universidad de
  Concepci\'on, Casilla 160-C, Concepci\'on, Chile}
\altaffiltext{4}{Millennium Institute of Astrophysics, Chile}
\altaffiltext{5}{Instituto de Astronom\'ia, Universidad Cat\'olica del
  Norte, Av. Angamos 0610, Antofagasta, Chile}

\begin{abstract}
By exploiting the exceptional high-resolution capabilities of the
near-IR camera GSAOI combined with the multi-conjugate adaptive optics
system GeMS at the GEMINI South Telescope, we investigated the
structural and physical properties of the heavily obscured globular
cluster Liller 1 in the Galactic bulge. We have obtained the deepest
and most accurate color-magnitude diagram published so far for this
cluster, reaching $K_s \sim 19$ (below the main sequence turn-off
level). We used these data to re-determine the center of gravity
  of the system, finding that it is located about $2.2\arcsec$
  south-east from the literature value. We also built new star
density and surface brightness profiles for the cluster, and 
  re-derived its main structural and physical parameters (scale
radii, concentration parameter, central mass density, total mass). We
find that Liller 1 is significantly less concentrated (concentration parameter $c=1.74$) 
and less extended (tidal
  radius $r_t=298\arcsec$ and core radius $r_c=5.39\arcsec$) than
previously thought. 
By using these newly determined structural parameters we estimated the mass of Liller~1
$M_{\rm tot} = 2.3~^{-0.1}_{+0.3} \times 10^6 M_\odot$ ($M_{\rm tot} = 1.5~^{-0.1}_{+0.2} \times 10^6 M_\odot$ for a Kroupa IMF), which is comparable
to that of the most massive clusters in the Galaxy ($\omega$ Centauri and Terzan~5).
Also Liller~1 has the second highest collision rate
(after Terzan 5) among all star clusters in the
Galaxy, thus confirming that it is an ideal environment for the
formation of collisional objects (such as millisecond pulsars).
\end{abstract}

\date{04 May 2015}


\section{Introduction}
Liller 1 is one of the 7 globular clusters (GCs) located within 1
  kpc from the Galactic centre, at a Galactocentric distance of only
  0.8 kpc, and very close to the galactic plane ($l = 354.84$, $b =
  -0.16$; \citealt{Ha96}, 2010 edition, hereafter H96). This region
of the Galaxy is strongly affected by a large foreground extinction: 
the average color excess $E(B-V)$ estimated by different authors in the direction of the cluster 
ranges from about 3.00 to 3.09 (\citet{Fr95}; \citealt{Or96,Or01}; \citet{Va10}) 
with significant evidence of differential
reddening \citep{Or01}. Such a large value of the extinction has
severely hampered the observations of Liller 1 in the optical bands.
Indeed very few papers about this cluster can be found in the
literature: the only available optical color-magnitude diagram (CMD) of Liller~1 is the one published by \citet{Or96} 
by using ESO-NTT data. More in general, the first CMD ever obtained for Liller~1 is the (K,J-K) CMD
published by \citet{Fr95}. However in both cases data are very shallow and the authors were able to sample only 
the brightest part of the red giant branch (RGB). The deepest photometry of Liller 1 published so far, 
was obtained by \citet{Da00} and \citet{Or01} using the
Canada-France-Hawaii Telescope and the Hubble Space Telescope, 
respectively. However, these samples are not deep enough to
properly characterize the main sequence turn-off (MS-TO) region of the
cluster. A more recent photometric analysis of Liller~1 was performed by \citet{Va10}
by using the large near-IR wide field of view (FoV) imager SofI mounted at the ESO-NTT.
The obtained CMD samples the
brightest portion of the RGB reaching the cluster red clump level
($K_s \sim 14$). 

Liller 1 is one of the most metal-rich GC in the Galactic bulge. In
fact by using high-resolution IR spectroscopy \citep{Or97}, both
\citet{Or02} and \citet{St04} measured a metallicity of about one half
solar ([Fe/H]$=-0.3$ dex). A similar value ([Fe/H]$=-0.36$ dex) was
also estimated by \citet{Va10} from a set of photometric indices (see
\citealt{Fe00}) characterizing the location and the morphology of the
RGB in the ($K_s, J-K_s$) CMD, and calibrated as a function of the metallicity
\citep{Va04a,Va04b}. 

One of the most astonishing characteristics of Liller 1 is the
extremely large value of the collision rate parameter.  \citet{VH87}
showed that Liller 1 has the second highest value of stellar encounter
rate (after Terzan~5; see also \citealt{La10}) among all star clusters
in the Galaxy, thus suggesting that it represents an ideal
environment where exotic objects, generated by collisions, can
form. In fact, it is commonly believed that dynamical interactions in
GCs facilitate the formation of close binary systems and exotic
objects like Cataclysmic Variables (CVs), Low Mass X-ray Binaries
(LMXBs), Millisecond Pulsars (MSPs) and Blue Straggler Stars (BSSs)
\citep[e.g.][]{Ba92,Pa92,Fe01,Fe09a,Fe12,Ra05,PH06}. Moreover,
\citet{Hu10} found that clusters with large collisional parameters and
high metallicity \citep[see also][]{bella95} usually host more MSPs. 
Indeed Terzan 5\footnote{Note
  that this stellar system is suspected to not be a genuine GC,
  because it harbors at least three stellar populations with different
  iron abundances \citep{Fe09b,Or11,or13,Ma14}.} hosts the largest
population of MSPs among all Galactic GCs \citep{Ra05}.  A strong
$\gamma$-ray emission has been recently detected in the direction of
Liller 1 by the Large Area Telescope (LAT) on board the Fermi
Telescope \citep{Ta11}. This is the most intense emission detected so
far from a Galactic GC, again suggesting the presence of a large
number of MSPs. However, no direct radio detection of these objects
has been obtained so far in this system \citep{Ra05}.  The only exotic
object identified in the cluster is the rapid burster MXB 1730-335, an
LMXB observed to emit radio waves and type I and type II X-ray bursts
\citep{Ho78}. It seems to be located in the central region of Liller
1, but no optical/IR counterpart of this object has been found so far
\citep{Ho01}.

In this paper we present the deepest near-IR photometry of Liller 1 yet obtained
and new determinations of its main structural and physical parameters.
In Section \ref{obs} we discuss the observations and data analysis.
Section \ref{resu} presents the results obtained, including new
determinations of: the distance modulus and reddening
(Sect. \ref{dist}), the center of gravity (Sect. \ref{centro}), the star density 
profile, the structural parameters and the surface brightness
(SB) profile (Sect. \ref{struct_param}), the total luminosity, total
mass and central mass density of the cluster (Sect. \ref{mass}).  The
summary and conclusions are presented in Section \ref{concl}.

\section{Observations and data analysis}
\label{obs}
\begin{figure}
\epsscale{.80}
\plotone{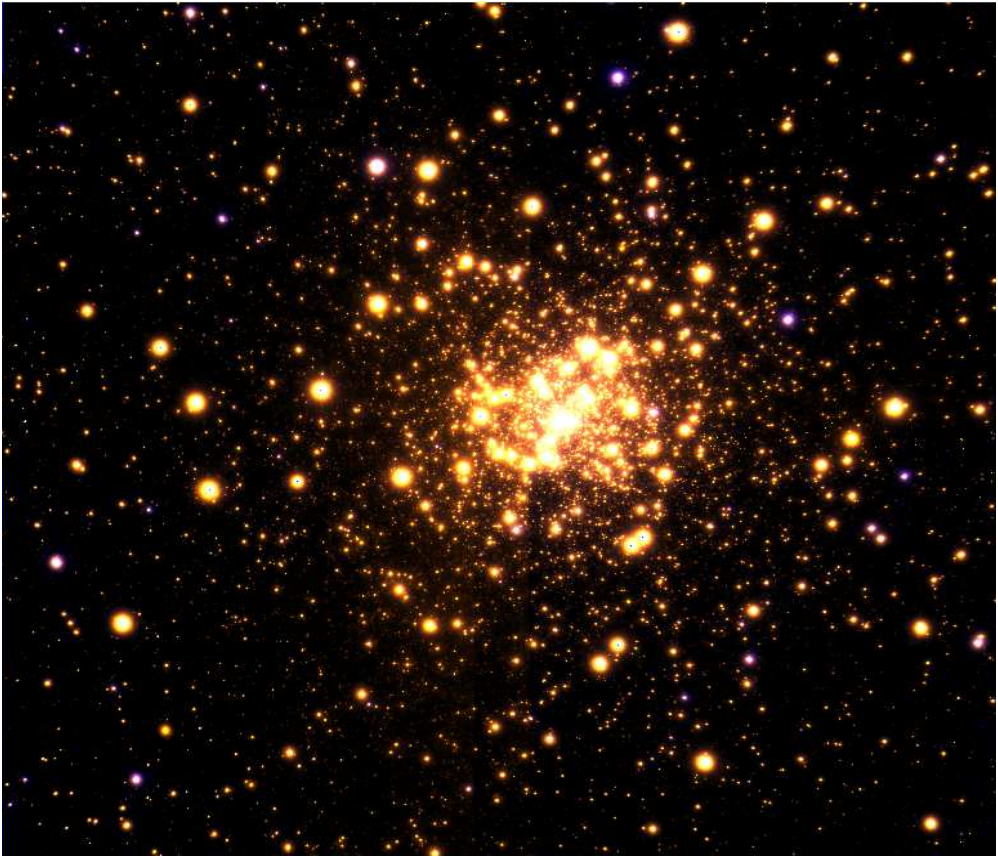}
\caption{False-color image of Liller 1 obtained by combining GEMINI
  observations in the near-IR $J$ and $K_s$ bands. North is up, east
  is on the right. The field of view is $85\arcsec\times
  85\arcsec$. The bluish stars are hot field objects.}
\label{fig1}
\end{figure}

The main photometric data used in the present study consist of a set of
high-resolution images obtained with the IR camera Gemini South
Adaptive Optics Imager (GSAOI) assisted by the Gemini Multi-Conjugate
Adaptive Optics System (GeMS) mounted at the 8 m Gemini South
Telescope (Chile), in April 2013 (Program ID: GS-2013-Q-23; PI:
D. Geisler).  GSAOI is equipped with a $2\times2$ mosaic of Rockwell
HAWAII-2RG $2048\times 2048$ pixels arrays. It covers a global FoV of $85\arcsec \times 85\arcsec$ on the sky, with a resolution
of $0.02\arcsec$/pixel \citep{Ne14}.  We sampled the central region of
Liller 1 with a mosaic of multiple exposures acquired with a dithering
pattern of a few arcseconds.  Seven and ten exposures have been
acquired in the $J$ and $K_s$ bands, respectively, with an exposure
time $t_{\rm exp}= 30$ s each. The entire data set was acquired in an interval of time of $\sim90$ mins.
A constellation of five laser guide
stars combined with three natural guide stars has been adopted to
compensate the distortions due to the turbulence of Earth's
atmosphere. We have verified that the Full Width at Half Maximum (FWHM) is quite stable allover the FoV
with a maximum variation of about $10-15\%$, in agreement with what found by \citet{Ne14}.
The average FWHM (i.e. estimated on the entire FoV) varies from $0.09\arcsec$ to $0.13\arcsec$ for the images acquired in the $J$ band and from $0.08\arcsec$ to $0.12\arcsec$ for the $K_s$ ones. These values are only slightly larger than the
diffraction limit of the telescope. In Figure \ref{fig1} we show a
two-color image of \objectname{Liller 1}, obtained by combining GEMINI
$J$ and $K_s$ band observations.

As discussed in Section \ref{resu} below, in order to derive the
radial star density and SB profiles of \objectname{Liller 1}, we
complemented this data set with a near-IR catalog  
from the  {\it VISTA Variables in the V\'ia L\'actea 
Survey} (hereafter VVV, \citealt{minniti10}; \citealt{catelan11})
complemented with the Two Micron All Sky Survey (2MASS; \citealt{Cu03}).

For the GEMINI data set, a standard pre-reduction procedure, using
IRAF\footnote{IRAF is distributed by the National Optical Astronomy
  Observatory, which is operated by the Association of Universities
  for Research in Astronomy, Inc., under cooperative agreement with
  the National Science Foundation.} tools, was applied to the raw
images to correct for flat fields and bias, and to perform the
sky-subtraction. The photometric reduction was then carried out via
point-spread function (PSF) fitting technique in each chip of each
image independently, using DAOPHOT
\citep{St87}, a package developed to perform accurate photometry in
crowded fields.  The PSF has been modeled by selecting about one
hundred bright and isolated stars uniformly distributed in each chip,
and by using the DAOPHOT/PSF routine.  We allowed the PSF to vary
within each chip following a cubic polynomial spatial variation. The
best-fit PSF analytic models obtained for the $J$ and $K_s$ images,
selected on the basis of a $\chi^2$ test, are a Penny function
\citep{Pe76} and a Moffat function with $\beta = 1.5$ \citep{Mo69},
respectively.  The PSF models thus obtained were then applied to all
the star-like sources detected at a 3$\sigma$ level from the local
background by using ALLSTAR. We have derived in this way the stellar
instrumental magnitudes. Then, starting from the star lists thus
obtained and to fill the gaps among the GSAOI chips, we created a
master star list containing all the stars measured in at least one $J$
and $K_s$ image.

As done in other works (see, e.g., \citealt{Da14}, and reference
therein), the master-lists thus created have been used as input for
ALLFRAME \citep{St94}. The files obtained as output have been combined
to have a complete catalog with the $J$ and $K_s$ magnitudes and the
positions of all detected stars. For every stellar source, different
magnitude estimates have been homogenized and their mean values and
standard deviations have been finally adopted as the star magnitudes
and photometric errors in the final catalog (see \citealt{Fe91,Fe92}).

\begin{figure}
\epsscale{.80}
\plotone{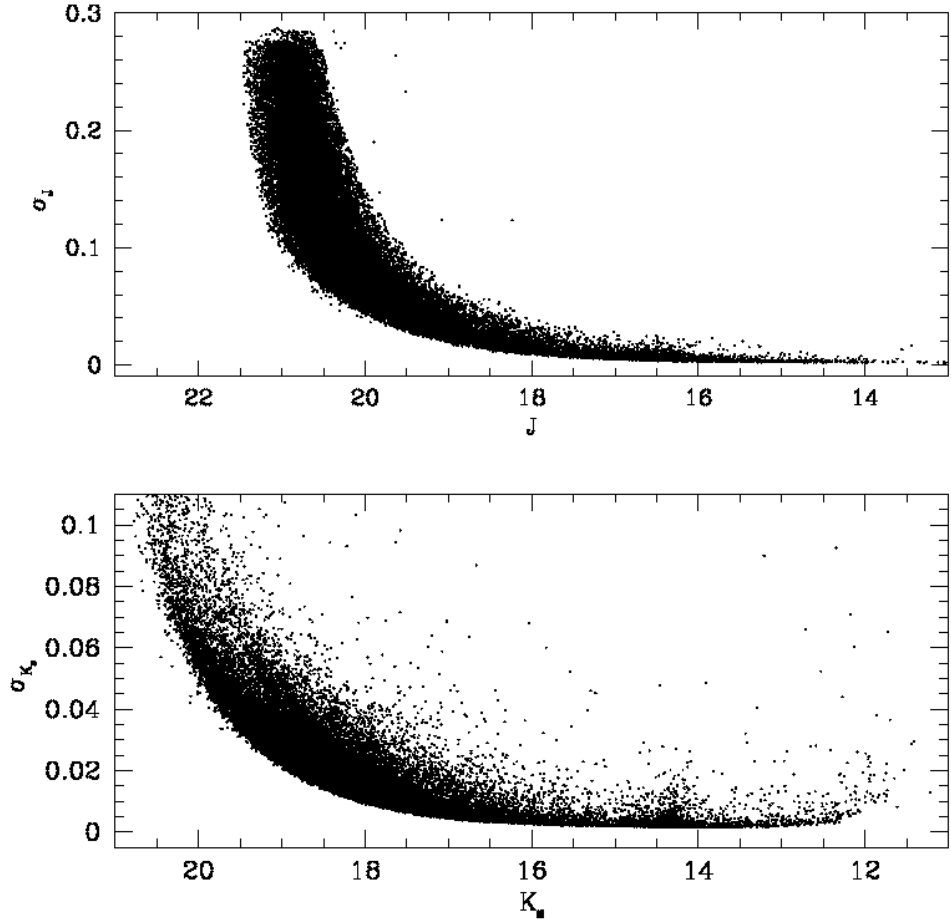}
\caption{Distribution of the photometric errors of the GEMINI data set as a function of $J$ and $K_s$ magnitudes (upper and lower panel, respectively).}
\label{errori}
\end{figure}

The distribution of the photometric errors as a function of the 
$J$ and $K_s$ magnitudes are shown in Figure \ref{errori}. 
Errors vary from 0.01 - 0.02 for bright RGB stars, and they reach 
values $\sigma_{rm K}\sim0.1$ and $\sigma_{rm J}\sim0.3$.
As expected, given the overall stability of the PSF, we do not observe any significant radial trend.

\begin{figure}
\epsscale{.80}
\plotone{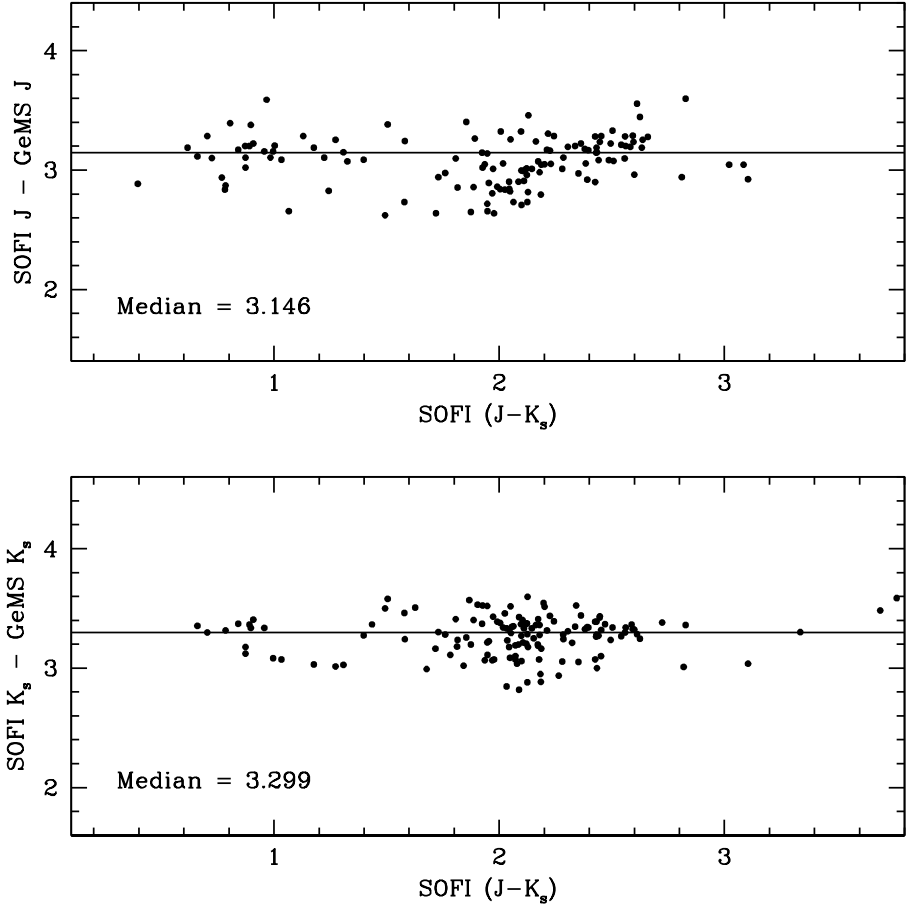}
\caption{Distributions of the (SOFI - GeMS) magnitudes as a function of the 
SOFI color $(J-K_s)$.
Stars at a distance larger than $30\arcsec$ from the cluster center and with 2.8 $<$ (SOFI $K_s$ - GeMS $K_s$) $<$ 3.6 and 2.6 $<$ (SOFI J - GeMS J) $<$ 3.6 respectively, are shown in figure.
The solid lines instead represent the median values obtained by applying an iterative $\sigma$-clipping procedure.
Finally we obtain (SOFI $K_s$ - GeMS $K_s$) = 3.299  and (SOFI $J$ - GeMS $J$) = 3.146.}
\label{calib}
\end{figure}

The instrumental magnitudes have been converted into the 2MASS
photometric system by using the stars in common with the publicly available 
catalog obtained with SOFI \citep{Va10}. 
To minimize the effect of blending, photometric errors and saturation, we 
used only stars at a distance larger than $30\arcsec$ from the cluster center and with 2.8 $<$ (SOFI $K_s$ - GeMS $K_s$) $<$ 3.6 and 2.6 $<$ (SOFI J - GeMS J) $<$ 3.6, respectively. The $J$ and $K_{s}$ calibration curves are shown in Figure \ref{calib}. 
As can be seen, we used an iterative sigma-clipping procedure to estimate the median of the (SOFI $K_s$ - GeMS $K_s$) and 
(SOFI $J$ - GeMS $J$) distributions for the stars satisfying the criteria listed above.
The r.m.s of the best-fit relation is 0.1 and 0.08 in $K_s$ and $J$, respectively. The values thus obtained (Figure \ref{calib}) have been applied to the instrumental GeMS magnitudes.
The same stars have been used also to put the instrumental positions 
onto the absolute astrometric system.

We retrieved VVV images obtained in the direction of Liller 1 
from the Vista Science Archive website\footnote{ http://horus.roe.ac.uk/vsa/}.
The data set is composed of one exposure per filter and it covers 
a total FoV of $33\farcm5\times33\farcm5$. 
Data were prereduced at the Cambridge Astronomical Survey Unit 
(CASU)\footnote{ http://casu.ast.cam.ac.uk/} with the VIRCAM pipeline \citep{Ir04}.
We then performed PSF-fitting photometry by using the DAOPHOT based 
VVV-SkZ\_pipeline \citep{mauro13} 
on the single $2048\times2048$ pixels chip extracted from the stacked VVV pawprints \citep{saito12}.
A quadratically variable Moffat function with $\beta = 3.5$ has been adopted.  
The VVV magnitudes and instrumental coordinates were then reported to the 
2MASS photometric and astrometric system \citep[see details in][]{chene12,monibidin11,mauro13}.

We estimated the differential reddening in the direction of Liller 1 by adopting a
procedure similar to that discussed in \citet{Ma12}, with the only difference that here
we used the RGB instead of the MS stars as reference sequence. In particular, we
divided the GSAOI FoV in a grid of $6\times6$ cells, each 18\arcsec-wide. We considered
RGB stars approximately in the magnitude range $14.5 < K_s < 16.5$: the upper and the lower
thresholds of the selection box have been set running parallel to the reddening vector. 
Considering the RGB stars within each cell, we estimated the  median
color.  The cell with the bluest color (nominally corresponding to the 
lowest extinction value) has been adopted as reference. The relative color shift $\delta [(J-K_s)]_i$
of each {\it i}th cell, is then defined as the shift needed to make 
the median color of
the {\it i}th cell match the reference cell color.  Then, from the value of $\delta
[(J-K_s)]_i$ we derived the corresponding $\delta [E(B-V)]_i$ by adopting the extinction
coefficients  R$_{\rm J}=0.87$ and R$_{\rm K_s}=0.35$ \citep{C89}. 
Extinction variations as large as $\delta [E(B-V)]=0.34$ mag have 
been measured in the direction of the cluster. 

\begin{figure}
\epsscale{.80}
\plotone{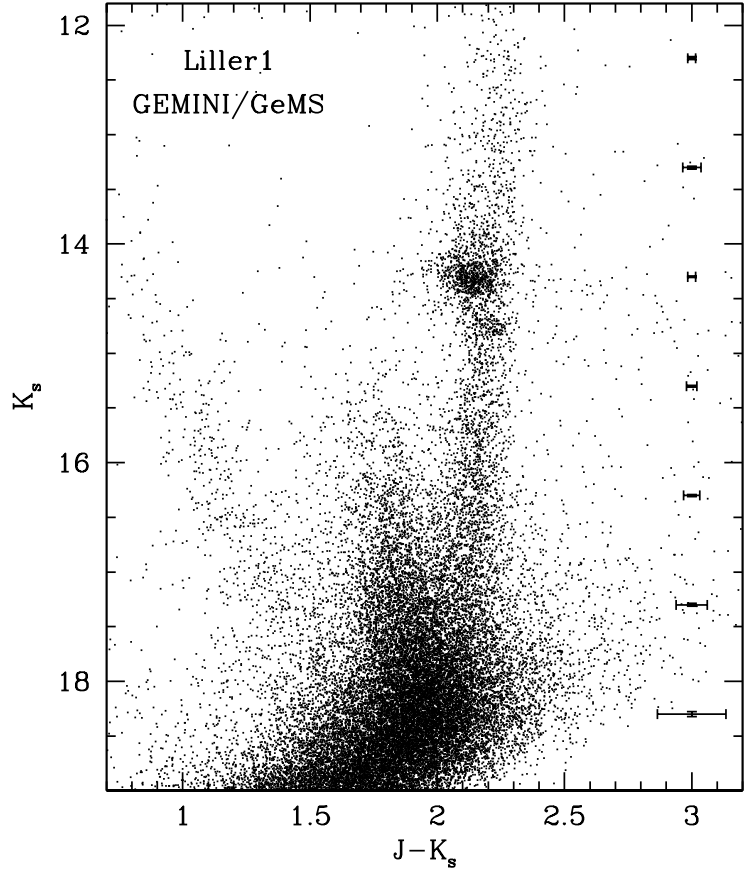}
\caption{Near-IR differential reddening-corrected CMD of Liller 1
  obtained from the GEMINI observations discussed in the paper. The
  main evolutionary sequences of the cluster are well visible down to
  the MS-TO point.  On the blue side of the CMD, for $(J-K_s) < 1.5$,
  the blue plume defined by the Galactic field MS is also
  distinguishable. The photometric errors for each bin of $K_s$
  magnitudes are shown on the right side of the panel.}
\label{fig2}
\end{figure}

\begin{figure}
\epsscale{.80}
\plotone{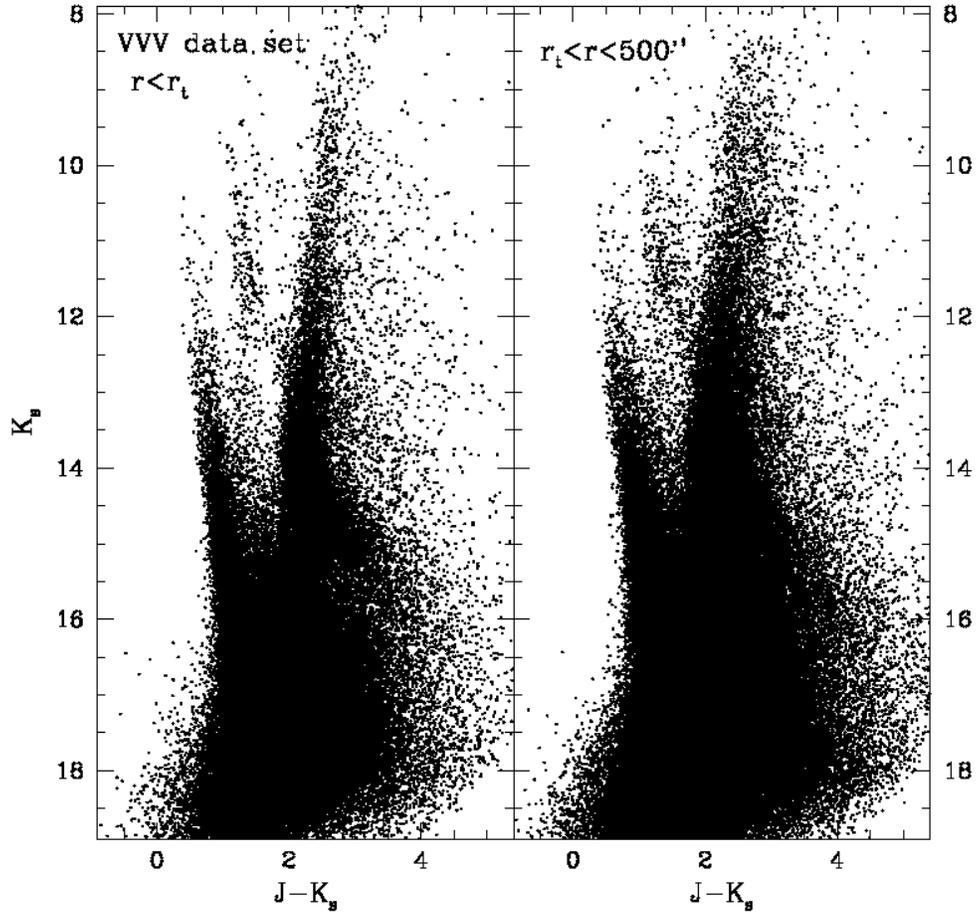}
\caption{Near-IR CMDs of Liller~1 as obtained from the VVV data. 
In the {\it left panel} only stars external to the GEMINI FoV and located at distances smaller than $r_t$ from $C_{\rm
grav}$  are shown. In the {\it right panel} only stars located at $r_t<r<500\arcsec$.  }
\label{VVV}
\end{figure}

Figure
\ref{fig2} shows the differential reddening corrected ($K_s, J-K_s$) CMD of
\objectname{Liller 1}. This is the deepest and most accurate CMD ever obtained for this
stellar system, reaching $K_s \sim 19$ and thus sampling the MS-TO region. Its well
defined red clump, clearly visible at $K_s\sim 14.2$, is a typical feature of a metal rich GC. 
Unfortunately, stars lying along the brightest portion of the RGB ($K_s < 12$) are
saturated in all the available images. As apparent from Figure \ref{VVV}, 
\objectname{Liller 1} suffers from significant field contamination mainly from the
Galactic bulge and disk stars, which define a blue plume clearly visible in the bluest
portion of the CMD for $(J-K_s) < 1.5$.
While the features of the CMD will be discussed in detail in a
forthcoming paper (hereafter, paper II), here we focus on the
derivation of the star density profile of Liller 1 and the estimate of
its main structural and physical parameters.

\section{Results}
\label{resu}
\subsection{Distance and reddening  of Liller 1}
\label{dist}
\begin{figure}
\epsscale{.80}
\plotone{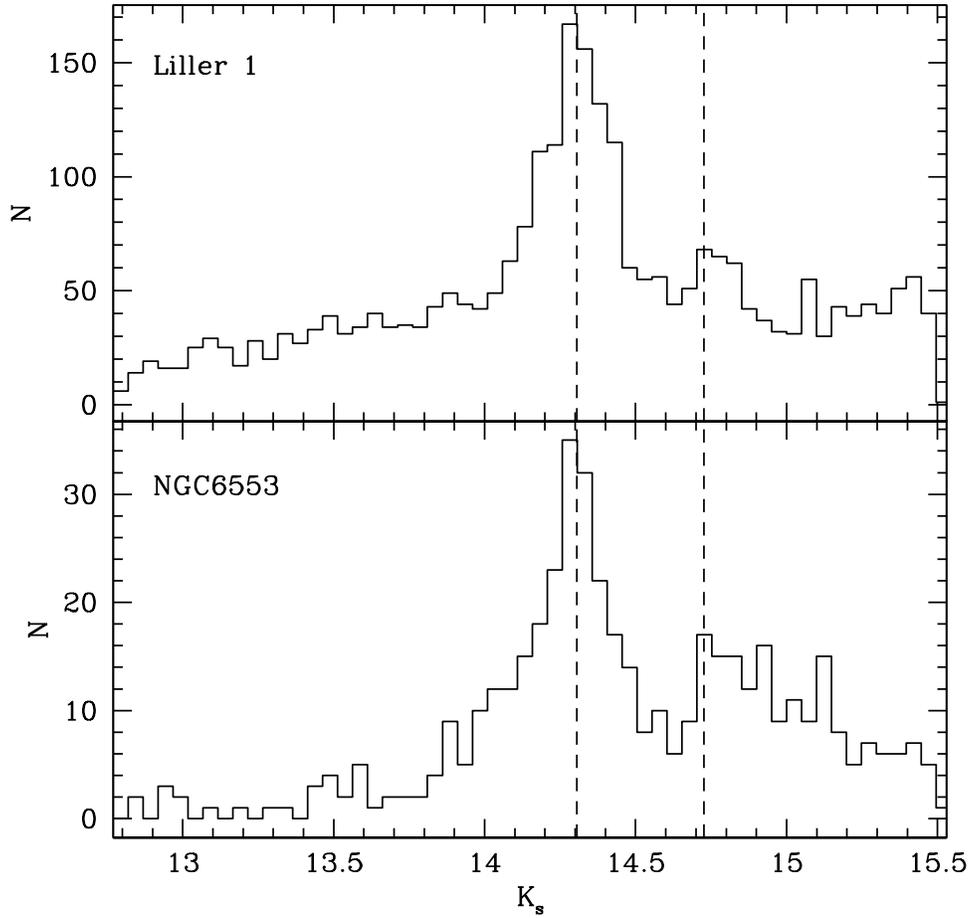}
\caption{Luminosity function of the red clump and the brightest
  portion of the RGB in Liller 1 (upper panel) and in the reference
  cluster NGC 6553 shifted by $\delta K=1.9$ (lower panel). The two
  dashed vertical lines mark the location of the red clump and the RGB
  bump.}
\label{fig:6553}
\end{figure}

The most recent determinations of the distance and reddening of Liller
1 \citep{Va10} have been obtained on the basis of a relatively shallow
$(K_s, J-K_s)$-CMD, reaching only the red clump level. Since the GEMINI
photometry presented here is significantly deeper, we used our new
data to re-determine both these quantities.  Following \citet[][see
  also \citealp{Da08a}]{Va07}, we used a differential method,
consisting in the comparison between the CMD and the luminosity
function of Liller 1 with those of a reference cluster, NGC 6553, a bulge GC with similar
metallicity \citep{Or02,melendez03,alves06,Va07}. The data set
  used for NGC 6553 is from \citet[][see
  also \citealp{Fe00}]{Va07}. We found that, in order to
align the RGB mean ridge line and the magnitude level of the red clump
of the two clusters, a color shift $\delta (J-K_s)=1.2$ and a magnitude
shift $\delta K_s=1.9$ must be applied to the sequences of NGC 6553 (see
Figure \ref{fig:6553}).  Thus, by adopting $(m-M)_0=13.46$ and
$E(B-V)=0.84$ for NGC 6553 \citep{Va10}, we obtain $(m-M)_K=15.65\pm0.15$ and
$E(B-V)=3.14\pm0.20$ for Liller 1.
The relative uncertanties on $(m-M)_K$ and $E(B-V)$ are obtained by taking into account
  the photometric errors
  at the magnitude level of the red clumps and RGB bumps of Liller~1 and NGC~6553, the errors due to the
  photometric calibration (Section \ref{obs}) and the histogram bin size.   
 
 It is important to note that the adopted reddening value has been obtained by
using the differential reddening-corrected CMD of Liller 1,
hence it corresponds to the least extincted region in the GEMINI FoV (see Section \ref{obs}). 
By averaging the 
values of E(B-V) of each cell considered in Section \ref{obs} 
weighted by the number of reference stars
sampled, we obtain a mean extinction  of $E(B-V)=3.30\pm0.20$ in the direction of Liller~1.
This yields a true (unreddened) distance modulus of $(m-M)_0=14.55 \pm 0.25$, 
which corresponds to a distance of $8.1\pm
1.0$ kpc, in agreement with previous determinations
(Harris 1996; Valenti et al. 2010; Ortolani et al. 2007).
 
Interestingly, Figure \ref{fig:6553} also shows that the relative
positions of the red clump and of the RGB bump
\citep{fusipecci90,Fe99a} are quite similar in the two
clusters. Hence, since the location in magnitude of the RGB bump is
quite sensitive to metal abundance, such a nice correspondence further
supports the evidence that Liller 1 and NGC 6553 share the same
chemistry.

\subsection{Center of gravity}
\label{centro}
Using the absolute positions of individual stars in the GEMINI sample,
we determined the center of gravity ($C_{\rm grav}$) of
\objectname{Liller 1}, by following the iterative procedure described
in \citet[][see also \citealt{Fe03,La07}]{Mo95}, averaging the right
ascension ($\alpha$) and declination ($\delta$) values of properly
selected stars.  As discussed in \citet[][see also
  \citealt{Lu95,Mi13}]{Fe03}, the use of individual stars to measure
the center of gravity and the star density profile provides the most
robust approach to derive cluster structural parameters, since the
counting of individual stars is not affected by the presence of a few
bright objects (which can, instead, generate spurious luminosity
clumping in the measurements based on the observed SB). However star
counts can be affected by incompleteness. We estimated the effect of completeness 
by means of artificial stars experiments. We followed the approach 
and prescriptions extensively described in \citet[][see also
  \citealt{B12}]{D11}. We obtained that only stars with $12.5 < K_s < 15.5$ have 
a completeness $C\sim100\%$ at any distance from the cluster center. 
We used only these stars to study both the center of gravity and the 
density profile (Section~\ref{struct_param}). 

We estimated the
gravity center by considering different sub-samples of stars: we used
stars lying within different distances ($10\arcsec$, $20\arcsec$ and
$30\arcsec$) from a first guess center, and with three different cuts in
magnitudes ($K_s$ =15.5, 15.0 and 14.5). The final value of $C_{\rm grav}$ is the average of the
different estimates. It turns out to be located at $\alpha_{\rm
  J2000}=17^{\rm h} 33^{\rm m} 24.56^{\rm s}$, $\delta_{\rm
  J2000}=-33^\circ 23\arcmin 22.4\arcsec$, with an uncertainty of
  $\pm0.3\arcsec$ and $\pm0.2\arcsec$ in $\alpha$ and $\delta$, respectively. Our center is $\simeq
2.2\arcsec$ south-east from that reported by H96 and estimated by
using the SB distribution.

\subsection{Star density profile and structural parameters}
\label{struct_param}
By adopting the derived value of $C_{\rm grav}$ and the data sets
described in Section \ref{obs}, we constructed the projected density profile
of the cluster over its entire radial extension (out to $900\arcsec$).  Following 
the procedure described in
\citet[][see also \citealt{La07,Da08b}]{Fe99b}, we divided the GEMINI and the VVV
samples in several concentric annuli centered on $C_{\rm
  grav}$, each one split into a variable number of sub-sectors. The
number of stars lying within each sub-sector was counted, and the star
density was obtained by dividing these values by the corresponding
sub-sector areas. The stellar density in each annulus was then
computed as the average of the sub-sector densities and the standard
deviation was adopted as the uncertainty.  In order to avoid
incompleteness and saturation biases, different limiting magnitudes
have been adopted for the two data sets: {\bf $12.5< K_s<15.5$} for the
GEMINI sample and $12< K_s<15.5$ for the VVV data set.  By guaranteeing an 
adequate radial overlap ($\Delta r=30-50\arcsec$) among the
two samples, the annuli in common between the adjacent data sets
have been used to join the two portions of the profile. At the end
of the procedure, the inner portion ($r<50\arcsec$) is sampled by the
high-resolution GEMINI data, while the outermost portion is obtained from the VVV data set.

\begin{figure}
\epsscale{.80}
\plotone{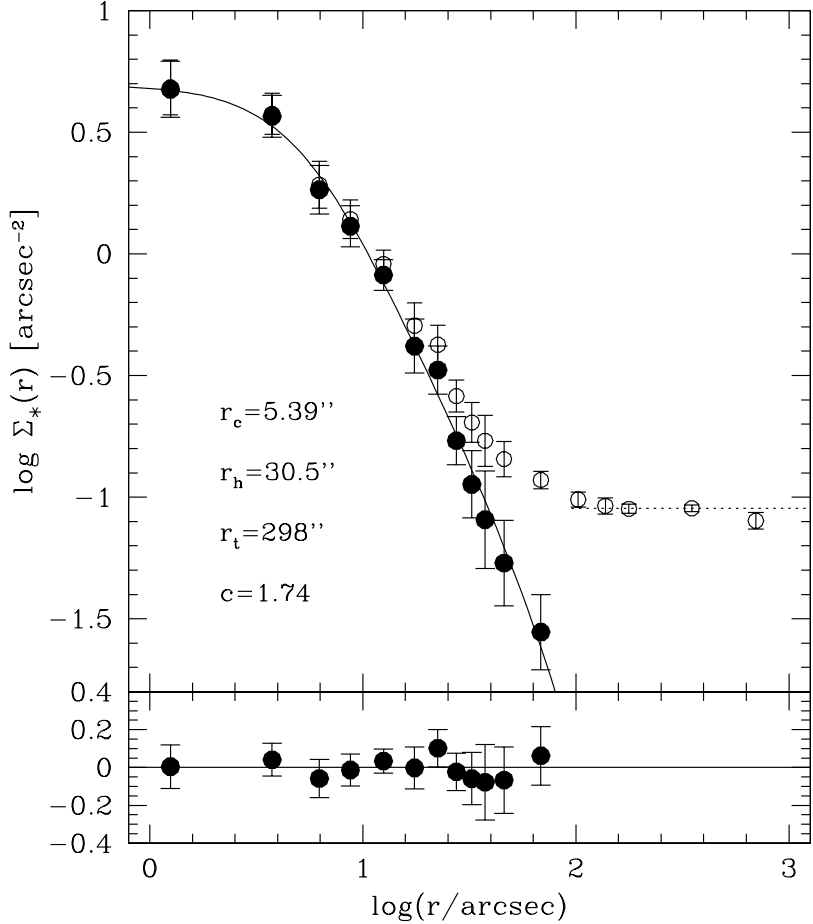}
\caption{Star density profile of Liller 1, obtained from resolved star
  counts in the combined data set. Empty circles represent the
  observed profile, while solid ones are obtained after subtraction of
  the Galactic field density (marked with the dotted line).  The
  best-fit single-mass King model is shown as a solid line and the
  corresponding structural parameters are labeled in the figure. The
  lower panel shows the residuals between the observations and the
  fitted profile at each radial coordinate.}
\label{fig3}
\end{figure}

The projected density profile thus obtained is shown in Figure
\ref{fig3} (empty circles). The distance from the center associated with
each point corresponds to the mid-value of each adopted radial bin.
The projected density at $r>100\arcsec$ is approximately constant,
consistent with being due to the Galactic field population alone.  To
estimate the background and foreground Galaxy contamination we
therefore used the $1\sigma$-clipped average of the five outermost
points (dotted line in Figure \ref{fig3}). The field-decontaminated
star density profile of Liller 1 is marked by the solid circles in the
figure.
The observed density distributions of GCs are traditionally
  described by means of King models \citep{Ki66}, even if deviations
  from this kind of profile have been found in some cases \citep[see,
    e.g.,][and references therein]{mclvdm05,monibidin11,Mi13}. To
  reproduce the observed star density profile of \objectname{Liller 1}
  and derive its structural parameters we used an isotropic,
  single-mass King model \citep{Ki66}, and we followed the procedure
  fully described in \citet{Mi13}.  According to a $\chi^2$ test, the
density profile of \objectname{Liller 1} can be excellently reproduced by a King
model with core radius $r_c = 5.39\arcsec^{-0.53}_{+0.61}$, concentration
parameter $c = 1.74^{-0.15}_{+0.15}$, half-mass radius $r_h =
30.5\arcsec^{-4.7}_{+7.9}$ and tidal (or limiting) radius $r_t =
298\arcsec^{-63}_{+82}$.

The structural parameters we obtained are significantly different from
those quoted by H96 based on the SB profiles in $I$ and $R$ bands
($r_c = 3.6\arcsec$, $c = 2.3$ and thus $r_t = 720\arcsec$). In
particular, the newly derived parameters show that Liller 1 is
significantly less concentrated and less extended than previously
thought. A reasonably good agreement is found with the $r_c$ values
estimated from the analysis of $JHK_s$ images 
by \citet[][$r_c=5.2\arcsec\pm 0.5\arcsec$]{Co07}\footnote{It is worth
  noticing, however, that the approach used in \citet{Co07} to fit the
  SB profile is quite different from ours, since the authors derived
  the central SB and the core radius of the cluster after adopting
  $r_t$ and $C_{\rm grav}$ from the online database of H96. Hence, the
  results obtained are not directly comparable.}  and by 
\citet[][$r_c=7\arcsec \pm 2\arcsec$]{Ma80}. A plausible
  explanation for the disagreement between the structural parameters
  derived from optical and near-IR observations could be a
  differential reddening effect, that may produce a spurious
  distortion of the optical SB profile, as well as low $S/N$ in the optical 
  due to the heavy extinction.

\begin{figure}
\epsscale{.80}
\plotone{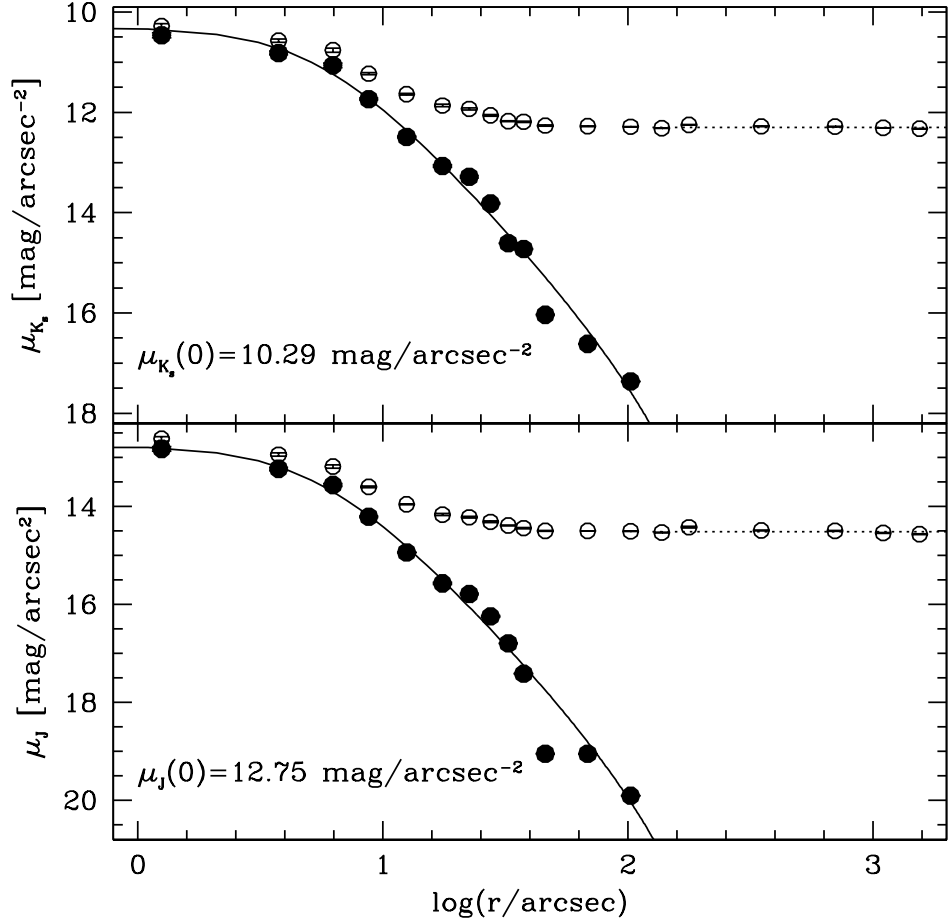}
\caption{SB profiles of Liller 1 in the $K_s$ and $J$ bands (top and
  bottom panels, respectively), obtained from the 2MASS data
  set. Empty circles represent the observed profile, while solid ones
  are obtained after background subtraction (marked with a dotted
  line). In both panels, the solid line corresponds to the best-fit
  King model shown in Figure \ref{fig3}.}
\label{fig4}
\end{figure}

We derived also the SB profiles of
Liller 1 directly from the images. To this aim, 
we used the wide-field images of the 2MASS data-set, since the VVV images heavily 
suffer from saturation.
We
produced $J$ and $K_s$ SB profiles by following the procedure
described in \citet[][see also \citealt{Co07}]{Da12} and by adopting the same center,
and a setup of annuli and sub-sectors similar to the one used 
to obtain the density profile. Basically, for
each annulus we summed the photon counts sampled in each sub-sector
and we assumed as final SB the average value of the different
sub-sectors.  In order to reduce statistical fluctuations due to the
presence of a few bright giants randomly distributed in the FoV 
of the $J$ and $K$ 2MASS images, a threshold of 8000 photon
counts was adopted for each pixel.  The counts have been converted
into instrumental magnitudes and then calibrated by using appropriate
zero-points. The contribution of the Galactic background was finally
estimated for $r>130\arcsec$, 
obtaining $\mu_{K_s}^{\rm back}= 12.3$ mag/arcsec$^2$ and
$\mu_J^{\rm back}= 14.5$ mag/arcsec$^2$.  The SB profiles thus
obtained are shown in Figure \ref{fig4}. These profiles are well fit
by the same King model derived from the projected density
distribution, thus confirming the accuracy of the structural
parameters derived in the present work.  The values of the central SB
measured in the $J$ and $K_s$ bands, $\mu_J(0)$ and $\mu_K(0)$, are
listed in Table \ref{tab1}, together with all the relevant parameters
derived for the cluster.

\subsection{Cluster physical parameters}
\label{mass}
Given the value of the effective radius estimated from the best-fit
King model discussed in the previous Section ($r_e=22.4\arcsec^{-3.3}_{+5.6}$), we
derived the integrated $K_s$ luminosity of the cluster within $r_e$.
By using aperture photometry on the 2MASS images and after subtracting
the background contribution determined as described in Section
\ref{struct_param}, we obtained $K_s(<r_e) = 4.30^{-0.19}_{+0.12}$.  Hence, by
definition of $r_e$ (i.e., the projected radius including half the
total integrated light), the total integrated magnitude of the cluster
is $K_s = 3.55^{-0.19}_{+0.12}$.  This value is significantly brighter than previous
estimates. In fact \citet{Co07} found $K_s(<50\arcsec)= 4.5$, while
our estimate over the same cluster region is 0.6 mag brighter, $K_s(<
50\arcsec) = 3.9$. We emphasize, however, that our estimate is a
direct measure of the integrated $K_s$ magnitude over the 2MASS image,
while that of \citet{Co07} depends on the assumed cluster parameters
(which are different from those derived here, as discussed in Section
\ref{struct_param}).
 
By adopting a color excess $E(B-V) = 3.30\pm0.20$, a distance modulus
$(m-M)_0 = 14.55\pm0.25$ (see Section \ref{dist}), and a bolometric
correction $BC_K = 2.3$ appropriate for a population of intrinsic
color $(J-K_s)_0 = 0.75$ \citep{Mo93}, we estimated a bolometric
luminosity of about $L_{\rm bol}(<r_e) = 3.45^{-0.10}_{+0.45} \times 10^5 L_\odot$,
corresponding to a total luminosity $L_{\rm bol}=6.9^{-0.2}_{+0.9} \times 10^5
L_\odot$. From this value, the total mass of Liller 1 can be estimated
by assuming a mass-to-light ratio $M/L_{\rm bol} = 3.35$
\citep{MT00,Ma03} appropriate for a Salpeter Initial Mass Function (IMF), 
thus obtaining $M_{\rm tot} =2.3^{-0.1}_{+0.3} \times 10^6
M_\odot$. We also estimated the mass of Liller~1 assuming $M/L_{\rm bol} = 2.19$ as 
obtained by \citet{Ma03} for a Kroupa IMF. We obtain in this case 
$M_{\rm tot} =1.5^{-0.1}_{+0.2} \times 10^6 M_\odot$.

Note that such a large value of the mass is similar to that extimated 
for $\omega$ Cen \citep{merritt97} and Terzan 5 \citep{Fe09a,La10},
and rankes Liller 1 in the high-mass tail of the distribution observed for 
GC-like stellar systems 
in the Milky Way.

Same values of the total mass are obtained from the total
integrated $K$-band magnitude (see above) and adopting mass-to-light
ratios $M/L_K = 1.56$ and $M/L_K = 1.03$ \citep{MT00, Ma03} for a Salpeter and a Kroupa IMF, respectively. 

We also used the available data to get a new estimate of the central
mass density of the cluster. To do so, we followed the same procedure
described for $V$-band data in \citet{Dj93}. However, in order to
minimize the effects due to the strong and differential reddening
affecting the system, we re-derived eq. (5) of \citet{Dj93} in the
appropriate form for $K_s$-band photometry. By adopting 3.28 as the
$K_s$-band magnitude of the Sun, we obtain:
\begin{equation}
\log I_{0,K_s} = 0.4 [24.852 - \mu_K(0)],
\end{equation}
where $I_{0,K_s}$ is the central projected luminosity density in units
of $L_\odot$ pc$^{-2}$. From eq. (4) of \citet{Dj93} and the
structural parameters quoted in Section \ref{struct_param}, we then
derived the central luminosity density of the system in the
$K$-band. By assuming $M/L_K = 1.56$ appropriate for a Salpeter IMF, we finally converted it into the
central mass density, $\rho_0 \simeq 7.2^{-1.0}_{+1.2}\times 10^6 M_\odot$
pc$^{-3}$. Using instead a Kroupa IMF which gives $M/L_K = 1.03$, we obtain 
$\rho_0 \simeq 4.8^{-1.0}_{+1.2}\times 10^6 M_\odot$
pc$^{-3}$.
These values are between 5 and 7 times larger than those obtained from the
$V$-band central luminosity density quoted in H96, by assuming $M/L_V
= 5.3$ \citep[from][]{MT00,Ma03}.

\section{Summary and conclusions}
\label{concl}
By using GeMS+GSAOI at the Gemini South Telescope, we have obtained
the deepest CMD of the Galactic GC Liller 1 published so far, properly
sampling even the innermost regions of the system (except
  for some saturated stars). This allowed us to obtain new estimates
of the cluster distance and reddening, which essentially confirm
previous determinations. Significant differences with respect to the
literature values, instead, have been found for the cluster center and
structural parameters: with respect to the values quoted in the H96
catalog, Liller 1 turns out to be located $\sim 2.2\arcsec$ south-east,
to be significantly less concentrated ($c=1.74$, instead of $c=2.3$),
to have a larger core radius ($r_c=5.39\arcsec$, instead of
$r_c=3.6\arcsec$), and thus to be less extended overall
($r_t=298\arcsec$, instead of $r_t=720\arcsec$). Also its total mass
$M_{\rm tot} = 2.3 \times 10^6 M_\odot$ and central mass density
$\rho_0=7.2 \times 10^6 M_\odot$ pc$^{-3}$ ($M_{\rm tot} = 1.5 \times 10^6 M_\odot$ 
and $\rho_0=4.8 \times 10^6 M_\odot$ pc$^{-3}$ for a Kroupa IMF) appear to be a factor of a
few larger that those derivable from the $V$-band parameters listed in
the H96 catalog.

From these new estimates, we also re-determined the collisional
parameter of the system, which can be expressed as $\Gamma \propto
\rho_0^{1.5} \times r_c^2$ for virialized and King model systems.  We
find that $\Gamma$ is $\sim$ 20 - 40 times larger than the one obtained by
using the H96 parameters. It corresponds to about half the collisional
parameter of Terzan 5 estimated in \citet{La10}, and is much larger
than the values derived for other massive GCs for which the structural
parameters have been re-determined consistently with what is done here
\citep[NGC 6388, NGC 6266, 47 Tucanae;
  see][respectively]{Da08a,Be06,Ma06}.

Hence, our analysis confirms the previous suggestion \citep{VH87} that
the collisional parameter of Liller 1 is the second highest (after
that of Terzan 5) among all Galactic GCs.  In stellar systems with
high values of $\Gamma$, large populations of ``collisional'' objects
(like LMXBs, MSPs, BSS and CVs), whose formation is promoted by
single/binary encounters, should not be unusual (also depending on the
availability of the progenitors).  In this respect, it is interesting
to note that, while Terzan 5 hosts the largest population of MSPs ever
observed in a GC \citep{Ra05}, no MSPs have been detected so far in
Liller 1, the only exotic object being in fact a LMXB \citep[][see
  also the Introduction]{Ho01}.  The lack of detection of large
populations of MSPs, LMXBs and CVs might be due to observational
biases. In fact Liller 1 is located in a region of the Galactic plane
where diffuse emission is very strong. In this respect, deeper MSP
searches, perhaps with the Square Kilometer Array, could help
understanding the real MSP content of the system. On the other hand,
the strong $\gamma$-ray emission (the most intense among all Galactic
GCs) detected in the direction of the cluster \citep{Ta11} may suggest
the possible presence of numerous, but still hidden, MSPs.  Another
possibility, emerging from the results presented in this work, is that
the $\gamma$-ray emission is not associated with Liller 1. In fact
\citet{Ta11} localized the position of the intense $\gamma$-ray source
at $\sim 7.5\arcmin$ from the cluster center. While this distance is
smaller than the tidal radius quoted in H96 ($r_t=720\arcsec =
12\arcmin$), it is larger than the cluster tidal radius obtained in
this work ($r_t \sim 300\arcsec = 5\arcmin$). Even by taking into
account the $\gamma$-ray position error ($\sim 2\arcmin$ for the
Fermi-LAT observations), the location of the $\gamma$-ray source is
only marginally compatible with the newly determined radial extension
of Liller 1, thus opening the possibility that such a strong emission
is not coming from this stellar system.

\acknowledgments This research is part of the project {\it Cosmic-Lab}
(http://www.cosmic-lab.eu) funded by the European Research Council
under contract ERC-2010-AdG-267675. D.G. gratefully acknowledges
support from the Chilean BASAL Centro de Excelencia en Astrof\'{i}sica
y Tecnolog\'{i}as Afines (CATA) grant PFB-06/2007. F.M. gratefully
acknowledges the support provided by Fondecyt for project
3140177. S.V. gratefully acknowledges the support provided by Fondecyt
reg. 1130721.  
This work made use of data products from the Two Micron All 
Sky Survey, which is a joint project of the University of 
Massachusetts and the Infrared Processing and 
Analysis Center/California Institute of Technology, 
funded by the National Aeronautics and Space 
Administration and the National Science Foundation.

\begin{table}
\begin{center}
\caption{New parameters for Liller 1.}
\label{tab1}
\footnotesize
\begin{tabular}{ll}
\\
\hline
\hline
Parameter & Derived Value\\
\hline

Center of gravity & $\alpha_{\rm J2000} = 17^{\rm h}33^{\rm m}24.56^{\rm s}$ \\
 & $\delta_{\rm J2000} = -33^{\circ} 23\arcmin 22.4\arcsec$ \\
Reddening& $E(B-V) = 3.30\pm0.20$ \\
Distance Modulus & $(m-M)_0$ = 14.55 $\pm$ 0.15\\
Distance & d = 8.1 $\pm$ 1.0 Kpc \\
Core radius & $r_c = 5.39\arcsec~^{-0.53}_{+0.61}= 0.21$ pc\\
Effective radius & $r_e = 22.4\arcsec~^{-3.3}_{+5.6} = 0.88$ pc \\
Half-mass radius & $r_h = 30.5\arcsec~^{-4.7}_{+7.9} = 1.20$ pc\\
Tidal radius & $r_t = 298\arcsec~^{-63}_{+82} =11.74$ pc\\
Concentration & $c = 1.74~^{-0.15}_{+0.15}$ \\
Total luminosity & $L_{\rm bol} = 6.9~^{-0.2}_{+0.9} \times 10^5 L_\odot$ \\
Total mass & $M_{\rm tot} = 2.3~^{-0.1}_{+0.3} \times 10^6 M_\odot$ (Salpeter IMF)\\
 & $M_{\rm tot} = 1.5~^{-0.1}_{+0.2} \times 10^6 M_\odot$ (Kroupa IMF)\\
Central mass density & $\rho_0 \simeq 7.2~^{-1.0}_{+1.2}\times 10^6 M_\odot$ (Salpeter IMF)\\
 & $\rho_0 \simeq 4.8~^{-1.0}_{+1.2}\times 10^6 M_\odot$ (Kroupa IMF)\\
Central $K_s$-band SB & $\mu_{K_s}(0) = 10.29\pm0.17$ mag arcsec$^{-2}$ \\
Central $J$-band SB & $\mu_J(0) = 12.75\pm0.17$ mag arcsec$^{-2}$\\

\hline
\end{tabular}
\end{center}
\end{table}


\clearpage

\end{document}